\documentclass[a4paper]{jpconf}
\usepackage{epsfig}
\usepackage{graphicx}

\begin{document}

\title{Geoneutrinos in Borexino}

\author{Marco G. Giammarchi and Lino Miramonti
\footnote{talk given by M.G. Giammarchi}}

\address{Dipartimento di Fisica dell'Universit\`a di Milano and Infn. \\ Via Celoria 16, 20133 Milano, Italy}

\begin{abstract}
This paper describes the Borexino detector and the high-radiopurity studies and tests that
are integral part of the Borexino technology and development. The application of Borexino
to the detection and studies of geoneutrinos is discussed.
\end{abstract}.

\section{The Gran Sasso National Laboratory}

The Gran Sasso National Laboratory (Laboratori Nazionali del Gran Sasso - LNGS), home of the Borexino experiment, is the world's largest underground laboratory. It is located in the center of Italy in the highway tunnel between Teramo and L'Aquila under the "Monte Aquila" (Gran Sasso mountain). The laboratory is financed and operated by the Italian National Institute for Nuclear Physics (Infn). 
Its total underground volume is about 180,000 $m^{3}$ with an area greater than 13500 $m^{2}$. It is composed of three main experimental halls (20 m high, 18 m wide and 100 m long). 
The overburden rock is on the average about 1,400 m, equivalent to 3,700 meters of water. The muon flux is reduced by about 6 orders of magnitude to a value of approximately 1.1 muons per square meter per hour, whereas the neutron flux is of the order of $3\times 10^{-6}$ neutrons per square centimeter per second with energies greater than 2.5 MeV.

The rock of the Gran Sasso mountain has a density of 2.71 $\pm$ 0.05 $g \cdot cm^{-3}$, and consists mainly of $CaCO_{3}$ and $MgCO_{3}$ \cite{Catalano1986}. 
The primordial radionuclide content of the rock of Hall C is 0.66 $\pm$ 0.14 ppm for $^{238}$U, 0.066 $\pm$ 0.025 ppm for the $^{232}$Th and 160 ppm for K \cite{Wulandari2004}. The radioactive content of the concrete employed as experimental hall liner is 1.05 $\pm$ 0.12 ppm for $^{238}$U and 0.656 $\pm$ 0.028 ppm for the $^{232}$Th \cite{Bellini1991}.

The LNGS hosts about 15 experiments of astroparticle physiscs such as neutrino research, double beta decay
physics, dark matter studies and nuclear astrophysics. Interdisciplinary studies (biology, geology) are
also conducted in the LNGS underground location. 

The Borexino detector is located in one of the big underground experimental halls, hall C.  

\

\section{The Borexino detector}

Borexino is a real time experiment whose main goal is to study the low energy (sub-MeV)
solar neutrinos, and in particular the 862 keV $^{7}$Be solar neutrino line, through the 
neutrino-electron elastic scattering reaction. The maximum energy of the recoiling
electron is 664 keV and the experimental design threshold is set at 250 keV \cite{Alimonti2002}. 

Borexino is an unsegmented scintillation detector featuring 300 tonnes of well shielded liquid 
ultra-pure scintillator viewed by 2200 photomultipliers (PMT). The detector core is a transparent 
spherical vessel (Nylon Sphere, 100$\mu$m thick), 8.5 m of diameter, filled with 300 tonnes 
of liquid scintillator and surrounded by 1,000 tonnes of high-purity buffer liquid. The 
scintillator mixture is PC (Pseudocumene) and PPO (1.5 g/l) as a fluor, while the buffer liquid 
will be PC alone (with the addition of DMP as light quencher). The photomultipliers are 
supported by a Stainless Steel Sphere, which also separates the inner part of the detector from 
the external shielding, provided by 2400 tonnes of pure water (water buffer), see figure 
\ref{f:Borexino}.

An additional containment vessel (Nylon film Radon barrier) is interposed between the 
Nylon Sphere and the photomultipliers, with the goal of reducing Radon diffusion 
towards the internal part of the detector.

The outer water shield is instrumented with 200 outward-pointing PMT's serving as a veto for penetrating muons, the only significant remaining cosmic ray background at the Gran Sasso depth. 

The innermost 2200 photomultipliers are divided into a set of 1800 PMT's equipped 
with light cones (so that they see light only from the Nylon Sphere region) and a set of 400 
PMT's without light cones, sensitive to light originated in the whole Stainless Steel Sphere 
volume. This design greatly increases the capability of the system to identify muons crossing 
the PC buffer (and not the scintillator).

The Borexino design is based on the concept of a graded shield of progressively lower intrinsic 
radioactivity as one approaches the sensitive volume of the detector; this culminates in the use 
of 200 tonnes of the low background scintillator to shield the 100 tonnes innermost Fiducial 
Volume. In these conditions, the ultimate background will be dominated by the intrinsic 
contamination of the scintillator, while all backgrounds from the construction materials and 
external shieldings will be negligible.

Borexino also features several external plants and purification systems conceived to purify the experimental fluids (water, nitrogen and scintillator) used by the experiment.

The main problem of a real time experiment with such a low energy threshold is the natural 
radioactivity which is present in any environment and in any material. For these reasons an 
intense R$\&$D program has been carried out in the last ten years to develop methods for 
selecting low radioactivity materials and/or purify them. An effort in this field has to be 
complemented by a comparably thorough research concerning detection and measurement of very 
low radioactivity levels. In this context four purification methods have been developed: distillation, water 
extraction, stripping with ultrapure N$_{2}$, solid gel column (Si gel, Al gel) adsorption. 

Significative results have been achieved by the Collaboration as for example: 10$^{-16}-10^{-17}$
(g of contaminants/g of material) for $^{232}$Th and $^{238}$U family and a few $\mu$Bq of 
Rn-222 in gases and liquids. In addition the organic solvent selected by the collaboration 
showed a $^{14}$C concentration clearly below 10$^{-17}$ in its ratio to $^{12}$C; this impurity 
is particularly important because it cannot be removed by chemical purification processes.

For the measurements of these ultralow radioactivity levels, dedicated methods were developed. 
In addition to small-scale techniques (Ge underground detectors in Rn-free environments, 
Inductively Coupled Plasma Mass Spectometer, high sensitivity Neutron Activation, Atomic 
Absorption Spectroscopy etc...\cite{Arpesella2002}) a prototype of the Borexino detector, 
the Counting Test Facility (CTF), has been constructed on purpose and operated in the
Hall C of LNGS. 

The radiopurities and sensitivities reached are summarized below and correspond to the lowest 
radioactivity levels obtained by the Borexino Collaboration, in preparation of the experiment:

\begin{itemize}

  \item Bulk material radiopurities of 10$^{-10}$ g/g for $^{238}$U and $^{232}$Th, 
$\sim$10$^{-5}$ for $^{nat}$K, few tenths of mBq/kg for $^{60}$Co, 
have been measured with Ge detectors in construction materials such as  
stainless steel, photomultipliers, metal and plastic gaskets, products for 
PMT sealing, etc...

  \item Radon emanations of 10 $\mu$Bq/m$^{2}$ from plastic materials, 0.1 mBq/m$^{3}$
for Rn-222 and 1 mBq/m$^{3}$ for Ra-226 in water, 
below 1 mBq/m$^{3}$ for the N$_{2}$ used for scintillator stripping.

  \item Radiopurity levels of a few times 10$^{-15}$ g/g $^{238}$U, $^{232}$Th and 
$^{40}$K have been reached with ICMPS in measuring the Borexino and CTF shielding 
water.
 
   \item Sensitivities of few ppt for $^{238}$U and $^{232}$Th concentrations have been obtained in the Nylon
Sphere material measurements. 
   
   \item The radiopurity of the scintillator itself was measured to be at the level of 
few 10$^{-16}$ g/g for $^{238}$U, $^{232}$Th and $\sim$10$^{-18}$ for
$^{14}$C/$^{12}$C in the Counting Test Facility.

   \item Bulk radiopurity levels of 
10$^{-13}-10^{-14}$ g/g for Au, Ba, Ce, Co, Cr, Cs, Ga, Hg, In, Mo, Rb;
less than few 10$^{-15}$ g/g for Cd, Sb, Ta, W; 10$^{-16}-10^{-17}$ g/g for
La, Lu, Re, Sc, Th; less than 1x10$^{-17}$ g/g for U, have been reached by means
of Neutron Activation followed by $\beta$-$\gamma$ delayed coincidence analysis
applied to the scintillator.
   
   \item Kr and Ar contamination in nitrogen at 0.005 ppm (for Ar) and 
0.06 ppt (for Kr) were obtained and measured with noble gas mass 
spectrometry.

\end{itemize}

These results represent a milestone in the development of the Borexino detector and 
technique. Several of these concepts were incorporated in the construction of
the high purity systems for the treatment of the most critical liquid,
the scintillator of the experiment.


\

\section{The Counting Test Facility}

The CTF description and its performance have been published
elsewhere \cite{Bel96,CTF-98,CTF-98A}. In this section we simply review the main features of
this detector.

The CTF consists of an external cylindrical water tank ($\oslash $11$\times $10 m; $\simeq $%
1,000 t of water) serving as passive shielding for 4.8 m$^3$ 
of liquid scintillator contained in an inner spherical vessel (Inner Vessel) of 2.1 m in 
diameter and observed by 100 PMT's. An additional nylon barrier against Radon convection and a muon veto system were installed in 1999. Figure \ref{f:CTF} shows a picture of the CTF detector.

The radio-purity level of the water is $\simeq 10^{-14}$ g/g (U, Th), $\simeq 10^{-10}$ g/g
($^{nat}$K) and $<$ 5 $\mu $Bq/{\it l} for $^{222}$Rn \cite {Bel96,CTF-98A,Bala96}.

The organic liquid scintillator has the same composition as in Borexino. The yield of emitted 
photons is $\simeq $10$^4$ per MeV of energy deposited and the fluorescence peak emission is 
located at 365 nm. The principal scintillator decay time is $\simeq $3.5 ns in a small
volume, while for large volume (because of absorbtion and re-emission) this value is 4.5--5.0 ns. 
The attenuation length is larger than $5$ m above 380 nm \cite{Alimo2000}. 

The purification of the scintillator is performed by recirculation from the Inner Vessel
through a Radon stripping tower, a water extraction unit, a Si-Gel column extraction unit, and a 
vacuum distillation unit. The  $^{232}$Th and $^{238}$U contaminations in the CTF liquid 
scintillator were found to be less than $(2$--$5)\cdot 10^{-16}$ g/g.

The Inner Vessel for the liquid scintillator containment is made of nylon with a thickness of 500 $\mu 
$m, with excellent optical clarity at 350-500 nm. The collection of scintillation light is 
ensured by 100 PMT's mounted to a 7 m diameter support structure inside the CTF tank. 

The photomultiplier tubes are 8 inches (Thorn EMI 9351, the same as for Borexino) made of low 
radioactivity Schott 8246 glass and characterized by high quantum efficiency (26\% at 420 nm), 
limited transit time spread ($\sigma $ = $1$ ns), good pulse height resolution for single
photoelectron pulses (Peak/Valley = $2.5$), low dark noise rate (0.5 kHz), low after pulse 
probability (2.5\%), and a gain of 10$^7$. 

The PMT's are equipped with light concentrators 57 cm long and with 50 cm diameter aperture.
The PMT system provides an overall 20\% optical coverage for events taking place inside the
Inner Vessel. 
The number of photoelectrons per MeV measured experimentally is (300 $\pm $ 30)/MeV on average.

The total background rate in the 250-800 keV energy range is about 0.3 counts/yr$\cdot $keV$\cdot $kg 
and appears to be dominated by external background from Radon in the shielding water 
($\approx $30 mBq/m$^3$ in the region surrounding the Inner Vessel). The internal background was measured to be
less than 0.01 counts/yr$\cdot $keV$\cdot $kg.


\

\section{The Counting Test Facility related publications}

Data collected with the Counting Test Facility have contributed significantly to the best limits
on quantities such as neutrino magnetic moment, electron lifetime, nucleon decays in invisible
channels, violation of the Pauli exclusion principle, production of heavy-neutrinos in the sun.

Concerning the study of the stability of the electron, the CTF data have been analyzed to search for the 256 keV line of the gamma emitted in the decay channel $e \rightarrow \gamma \nu$. Since we have found no signal, we established a limit on the electron lifetime of $\tau \geq 4 \cdot 10^{26}$ (90\% C.L.); this is still the best world limit for the electron decay in this channel \cite{Back2002}.

CTF data analysis has allowed the study of the neutrino magnetic moment, obtaining the limit of 
$\mu_{\nu} \leq 0.5 \cdot 10^{-10} \mu_{B}$, still a very competitive result \cite{Back2003}.

We have also investigated the possibility of heavy neutrinos $(M \geq m_{e})$ emitted in the $^8B$ reaction in the sun. Heavy neutrinos would decay to light neutrinos via the reaction 
$\nu_{H} \rightarrow \nu_{L} + e^{+} + e^{-}$. The analysis of the CTF energy spectrum  has allowed to 
significantly enlarge the excluded region of the parameter space with respect to previous experiments \cite{Back2003b}.

The stability of nucleons bounded in nuclei has been studied in the Counting Test Facility searching for
decays of
single nucleon or pair of nucleons into invisible channels. The limits are comparable to or improve the previously set world limits \cite{Back2003c}. Furthermore a search was made for non-Paulian transitions of nucleons from nuclear $1P$ shell to a filled $1S_{1/2}$ shell obtaining the best limit on the Pauli exclusion principle \cite{Back2004}.

Other studies have concerned the search for anti-neutrinos coming from the sun \cite{CTFantinu} and the
cosmogenic $^{11}$C underground production \cite{CTFcosmo}. 

\

\section{Geoneutrinos detection}

One of the possible application of a high mass well shielded scintillator detector such as Borexino
is the search for geoneutrinos, a new and very interesting subject which we will discuss in the 
remaining of this paper.

The conceptual foundations of Earth science rest on a variety of observables
as well as interior characteristics. One of the most important interior
parameter is the internally produced heat which is currently measured to 
be in the $\sim 60$ mW/m$^2$ range (or 30 TW when integrated on the planet 
surface).

Part of this energy flow is due to the presence of radioactive elements in the Earth
interior, mainly naturally occurring Uranium and Thorium chain elements and
potassium. Models of the Earth disperse about 50\% of the total U,Th in the
crust while leaving the remaining half to the mantle. Roughly speaking, the 35 km
thick continental crust contains a few ppm of U,Th while the much thinner
($\sim$ 6 km) oceanic crust has a typical concentration of $\sim$ 0.1 ppm.

Our goal is to measure this radiogenic heat by detecting neutrinos emitted
during the decays of the radioactive chains. For a given structure of 
naturally occurring radioactive
families, a measurement of antineutrino flux can be related to 
the U,Th family content of the Earth.

In the case of the $^{238}$U family (the $^{235}$U leftover can be neglected) one 
can globally represent the full decay chain as:

\begin{equation}
^{238}U \rightarrow \, ^{206}Pb + 8\, \alpha + 6\, e^- + 6\, {\overline \nu_e} 
+ 51.7\, MeV
\end{equation}

and therefore the detected number of antineutrinos is related to the number
of times this reaction has taken place.

The $^{232}$Th family presents a similar case:

\begin{equation}
^{232}Th \rightarrow \, ^{208}Pb + 6\, \alpha + 4\, e^- + 4\, {\overline \nu_e} 
+ 42.8\, MeV
\end{equation}

where again a definite number of antineutrinos is emitted and a well defined
energy is released.

Finally, antineutrinos are also emitted in the K terminations:

\begin{equation}
^{40}K \rightarrow \, ^{40}Ca + e^- + {\overline \nu_e} 
+ 1.32\, MeV 
\end{equation}

The ${\overline \nu}$ energy spectra produced by these three sources are 
plotted in fig. \ref{f:spectra}.

\


Our goal will be to detect antineutrinos emitted by these sources in order
to measure the rates of occurrence of the above reactions.

\subsection{Principle of geoneutrino detection}

While low energy neutrino detection offers formidable experimental challenges
due to backgrounds as explained above, antineutrinos were discovered in a
reaction that naturally affords a nice way to cope with the unwanted 
background events. The proposed detection reaction:

\begin{equation}
{\overline \nu_e} \, + \, p \rightarrow n \, + \, e^+
\end{equation}

is the inverse beta decay (or {\it Reines-Cowan}) reaction and 
generates a positron and a neutron in the final state. The positron gets 
quickly absorbed in ordinary matter, coupling with an electron and generating 
two gammas ($e^+e^- \rightarrow \gamma \gamma$) with a total energy release 
of 1.02 MeV/c$^2$.

The neutron, on the other hand, gets slowed down in the material and 
finally thermalizes to be absorbed as shown in fig. 
\ref{f:reinescowan}. The lifetime of this
process in the Borexino scintillator is of about 200 $\mu s$ and at the end
the neutron is absorbed by free protons in the scintillator: 

\begin{equation}
H (n,\gamma) D
\end{equation}

with the emission of a 2.2 MeV energy gamma.


From the detection viewpoint, this cascade of processes allows a very
favorable tag of energies and time. 

First of all, the energy of the two electromagnetic cascades are above 
$\sim 1$ MeV and 2.2 MeV respectively. Secondly, the time delay between the two 
events is a short one, very difficult to mimic by
a couple of accidental background events. Finally, these two events are
also subjected to a mild spatial condition of coincidence.

In summary, the tagging of an antineutrino event will be an 
$E>$1 MeV $e/\gamma$ event followed (in a narrow time window of, say, 0.5 ms)
by a 2.2 $\gamma$ event. The two events must lie within the typical (1 m) 
neutron diffusion length.

The ${\overline \nu_e} \, + \, p \rightarrow n \, + \, e^+$ detection 
reaction has a threshold of of 1.8 MeV which is determined by the mass
difference between neutron plus positron and the initial state proton.
Therefore this reaction has the drawback of not
being sensitive to the detection of K antineutrinos (see fig. \ref{f:spectra}).

The detector will reconstruct the antineutrino energy based on the 
observed positron kinetic energy. The visible energy of the event has
to take into account also the 1.02 MeV energy due to annihilation:

\begin{equation}
E(vis) = K(e^+) + 1.02 \, MeV
\end{equation}

which is plotted in fig. \ref{f:visible} for the case of the U and Th chain.


In turns, the energy of the antineutrino is shifted with respect to
$K(e^+)$ by the Q-value of the reaction:

\begin{equation}
E({\overline \nu}) = K(e^+) - 1.8 \, MeV
\end{equation}

So, the final relation between the visible energy and the 
${\overline \nu}$ energy is 

\begin{equation}
E(vis) = E({\overline \nu}) -0.78 \, MeV
\end{equation}

The observed kinetic energy spectrum $E(vis)$ will begin at 1.02 MeV
(the case when the positron kinetic energy is zero). 

Apart from the internal (radioactive) contamination of the scintillator, 
the background to the antineutrino signal can come 
in principle from a variety of sources, including atmospheric shower
particle decays ($\pi$, $\mu$, $K$), relic of past supernovas, 
non-standard $\nu \rightarrow {\overline \nu}$ oscillations in the
Sun, muon induced neutron production and ${\overline \nu_e}$ from
nuclear reactors.

It can be shown (see ref. \cite{Raghavan} and \cite{Calaprice}) that
the most significant of these external backgrounds is by far the 
term coming from nuclear reactors. 
 
\subsection{Background from nuclear reactors}

Nuclear power reactors produce energy by fission of heavy nuclei.
Since a super-heavy nucleus has a $\sim$30\% excess of neutrons over
protons, this excess will be transferred to the lighter fission
products which therefore will be beta-instable and produce
antineutrinos during their deexcitation. 

In order to produce background signal for the inverse-beta decay 
detection reaction the antineutrinos must have more than 1.8 MeV 
kinetic energy. The fissile nuclides featuring fission fragments of such
energy are $^{235}$U, $^{239}$Pu, $^{238}$U and $^{241}$Pu.

The number of fissions generated is typically calculated from the
reactor thermal power and the specific fission energy release.
In addition, some mild dependence over time is introduced by the
initial composition and time evolution of nuclear fuel at a 
specific reactor.

However, in spite of this parameter variability, antineutrino reactor
spectra are all very similar, producing a background that extends 
from our detection threshold and up to about 9 MeV.

For the case of Borexino, the relevant reactors are the ones 
situated in Europe, which are relatively far away (700-800 km)
from the Gran Sasso location (fig. \ref{f:reactlocation}). 
Neutrino oscillation effects will then be washed off in phase giving only
an overall reduction factor to about 60\% of the original reactors flux.


Fig. \ref{f:geoandreactor} shows both the the reactor background
spectrum and the U,Th signal to be searched for. The normalization 
of the spectra is made accordingly to the expected number of events from european
reactors (20 events/year) and the estimated geoneutrino signal
in Borexino (6 events/yr). More on this later on.


\subsection{Background from $^{210}$Pb}

One of the most important internal backgrounds for the study of 
geoneutrinos comes from the $^{210}$Pb content of the scintillating material.
This nuclide has a 22.3 yr half-life and can be introduced in the 
detecting volume either by natural bulk ($^{226}$Ra) contamination or 
(perhaps most importantly) through $^{222}$Rn diffusion.

$^{210}$Pb has an $\alpha$-emitting daughter ($^{210}$Po) which has a 
138 days half-life and can give rise to the reaction

\begin{equation}
^{13}C(\alpha ,n)^{16}O
\end{equation}

This originates a chain of events closely mimicking the ${\overline \nu}$ event signal with a first
first n-p scattering release (or the $\alpha$ release) followed by 
neutron capture. In fact this reaction was one of the backgrounds considered
in the first detection of geoneutrinos by the Kamland experiment \cite{Kamland}.

In the case of the Borexino experiment, the studies conducted in 
the Counting Test Facility allowed a careful measurement of this
background component. In particular $^{210}$Pb was studied during long
runs by detecting the $^{210}$Po alpha and by studying the total single
rate spectra (see fig. \ref{f:ctfit} for an example of such fits.)


These studies showed that the specific $^{210}$Pb background in Borexino
is limited to $\sim$ 20 $\mu$Bq/ton, a value much lower than the
one quoted in \cite{Kamland}. Therefore, this background will be
negligible in Borexino.

\section{Antineutrino signal and detection sensitivity}
Having demonstrated that the dominant background in Borexino will be
the one due to european reactors, we now proceed to the evaluation
of the terrestrial antineutrino signal. 

In order to do this, it is necessary to take into account the detailed
distribution of U,Th in the Earth's crust, particularly within 500 km
distance from the detector (a region that generates about 1/2 of the 
total signal).

For this evaluation we have exploited the Earth's model given in 
\cite{Mantovani} which gives U,Th distributions and detailed evluations
for specific detector locations. This evaluation, together with the
high efficiency (assumed 1) of the detection, leads to the prediction
of $\sim$6 events/yr (oscillation effects included) detected in Borexino, as shown in fig. \ref{f:geoandreactor}.
 
\section{Conclusion}

Terrestrial antineutrino detection requires shielded low background
detectors of high mass. Borexino, located at Gran Sasso Laboratory, will tackle
this fascinating subject armed with its low background capability
and its relatively long distance from nuclear reactors. 

We estimate that Borexino will detect $\sim$6 events/yr coming from
geoneutrinos with some background from nuclear reactors (fig. \ref{f:geoandreactor}).
The statistical significance of such a signal is predicted to be better than
30\% in 5 years of data taking.

\

\par\noindent
{\bf Acknowledgements}. We would like to thank Stephen Dye, John Learned and all the organizers of this conference 
for providing us with a stimulating scientific atmosphere in a wonderful setting.

\

\end{document}